\documentclass{ws-procs9x6}
\begin{document}
\title{ NEUTRINO OSCILLATIONS: HIERARCHY QUESTION}
\author{D.~J.~Ernst, B.~K.~Cogswell}
\address{Department of Physics and Astronomy, Vanderbilt University, Nashville, Tennessee, 37235, USA
 \\Department of Physics, Fisk University, Nashville, Tennessee,37208}
\author{H.~R.~Burroughs}
\address{Department of Physics and Astronomy, Vanderbilt University, Nashville, Tennessee, 37235, USA}
\author{J.~Escamilla-Roa}
\address{Departamento de F\'{\i}isica, Universidad Aut\'onoma del Estado de Hidalgo, Pachuca, Hidalgo, M\'exico}
\author{D.~L.~Latimer}
\address{Department of Physics, University of Puget Sound, Tacoma, Washington, 98416}
\begin{abstract}
The only experimentally observed phenomenon that lies outside the standard model of the electroweak interaction is neutrino 
oscillations. A way to try to unify the extensive neutrino oscillation data is to add a phenomenological mass term to
the Lagrangian that
is not diagonal in the flavor basis. The goal is then to understand the world's data in terms of the parameters of the mixing matrix 
and the differences between the squares of the
masses of the neutrinos. An outstanding question is what is the correct ordering of the masses, the hierarchy question. We point out a broken
symmetry relevant to this question, the symmetry of the simultaneous interchange of hierarchy and the sign of $\theta_{13}$. We first
present the results of an analysis of data that well determine the phenomenological parameters but are not sensitive to
the hierarchy. We find $\theta_{13} = 0.152\pm 0.014$, $\theta_{23} = 0.25^{+0.03}_{-0.05}~\pi$ and $\Delta_{32} = 2.45\pm
0.14 \times 10^{-3}$ eV$^2$, results consistent with others. We then include data that are sensitive to the hierarchy and the sign of 
$\theta_{13}$. We find, unlike
others, four isolated minimum in the $\chi^2$-space as predicted by the symmetry. Now that Daya Bay and RENO have determined $\theta_{13}$ 
to be surprisingly large, the Super-K
atmospheric data produce meaningful symmetry breaking such that the inverse hierarchy is preferred at the 97.2 \% level.
\end{abstract}

\bodymatter
\section{Introduction}
Neutrinos undergo the phenomenon of flavor oscillations; a neutrino with a particular flavor (electron, mu, or
tau) will change its flavor, a phenomenon not included in the standard model of the electroweak interaction. A
possible explanation of this behavior is neutrino oscillations. The neutrinos are assigned masses. The created
neutrino has definite flavor, but these flavor eigenstates are not the physical free particles, the mass
states. For three neutrinos, this leads to a phenomenology that is parameterized by three mixing angles, a CP violating phase, and 
two mass-squared differences, $\Delta_{ij}=:m_i^2-m_j^2$. A goal of the field is to unite the experimental data
through the determination of these parameters. 

\begin{figure}[b]
\begin{center}
\includegraphics*[width=3.5in]{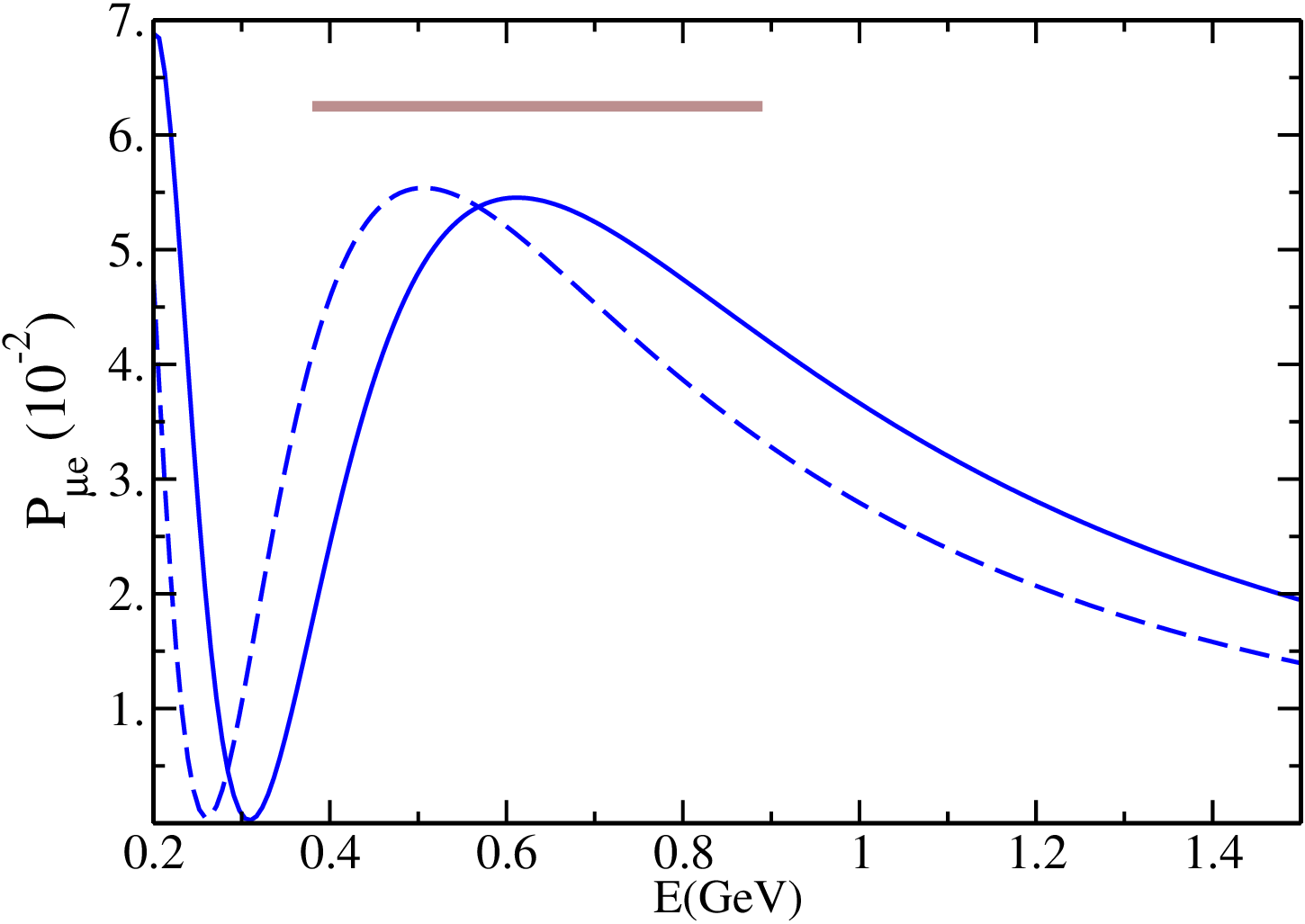}
\caption {${\mathcal P}_{\mu e}$  in vacuum versus neutrino energy for the T2K experiment. The solid curve depicts the normal 
hierarchy, positive $\theta_{13}$ as well as the inverse hierarchy, negative $\theta_{13}$ results. The dashed curve 
depicts the normal hierarchy, negative $\theta_{13}$ as well as  the inverse hierarchy, positive $\theta_{13}$ results. 
The horizontal bar represents the range where the muon interaction rate at T2K is greater than 50\% of its maximum. }
\label{fig1}
\end{center}
\end{figure}

Neutrino oscillations approximately break into two uncoupled
oscillations, solar and atmospheric oscillations. The solar oscillations are predominantly governed by the solar
data that determine the mixing angle $\theta_{12}$ and the KamLAND reactor data that determine the mass-squared
difference $\Delta_{21}$. We fix these two parameters to values determined by others. With them fixed, we perform a full
three-neutrino analysis. 

An outstanding question in neutrino phenomenology is that of hierarchy, whether the small mass-squared difference 
lies below (normal hierarchy) or above (inverse hierarchy) the large mass-squared difference. A related question, as 
we will show, is whether the value of $\theta_{13}$ is positive or negative. We utilize the convention\cite{Lat05} on
the bounds of $\theta_{13}$ of  
$-\frac{\pi}{2}\le\theta_{13}< +\frac{\pi}{2}$ and $0\le\delta<\pi$. These are the more convenient bounds for 
discussing the symmetry of interest here. Recent long-baseline reactor experiments \cite{Abe12,An12,Ahn12} have measured 
the value of $\theta_{13}$ and found it to be large, $\sin^2(2\,\theta_{13})\approx 0.1$. Before these measurements, the 
world's data demonstrated \cite{Esc10} very little preference for either the hierarchy or the sign of $\theta_{13}$. A goal 
of this work is to examine the effect of the knowledge of $\theta_{13}$ on the hierarchy question and the 
sign of $\theta_{13}$. This report is preliminary in that we make use of the atmospheric analysis of Ref.~\refcite{Roa09} by
including its constraints on $\theta_{23}$ and $\theta_{13}$ separately, ignoring correlations\cite{Lat05b}. No significant 
evidence for CP violation has been found\cite{For12,Fog12} so for now we set the CP phase $\delta$ to zero. 

In Section 2 we discuss the symmetry that is relevant for the investigation of the hierarchy question. In
Section 3 we present results of our analysis of data that do not have sensitivity to the hierarchy or the sign of
$\theta_{13}$. We then include data that are sensitive to the hierarchy and the sign of
$\theta_{13}$ in Section 4. We finish with a discussion of the significance of this work and point out
future  work.

\begin{figure}[t]
\begin{center}
\includegraphics*[width=3.5in]{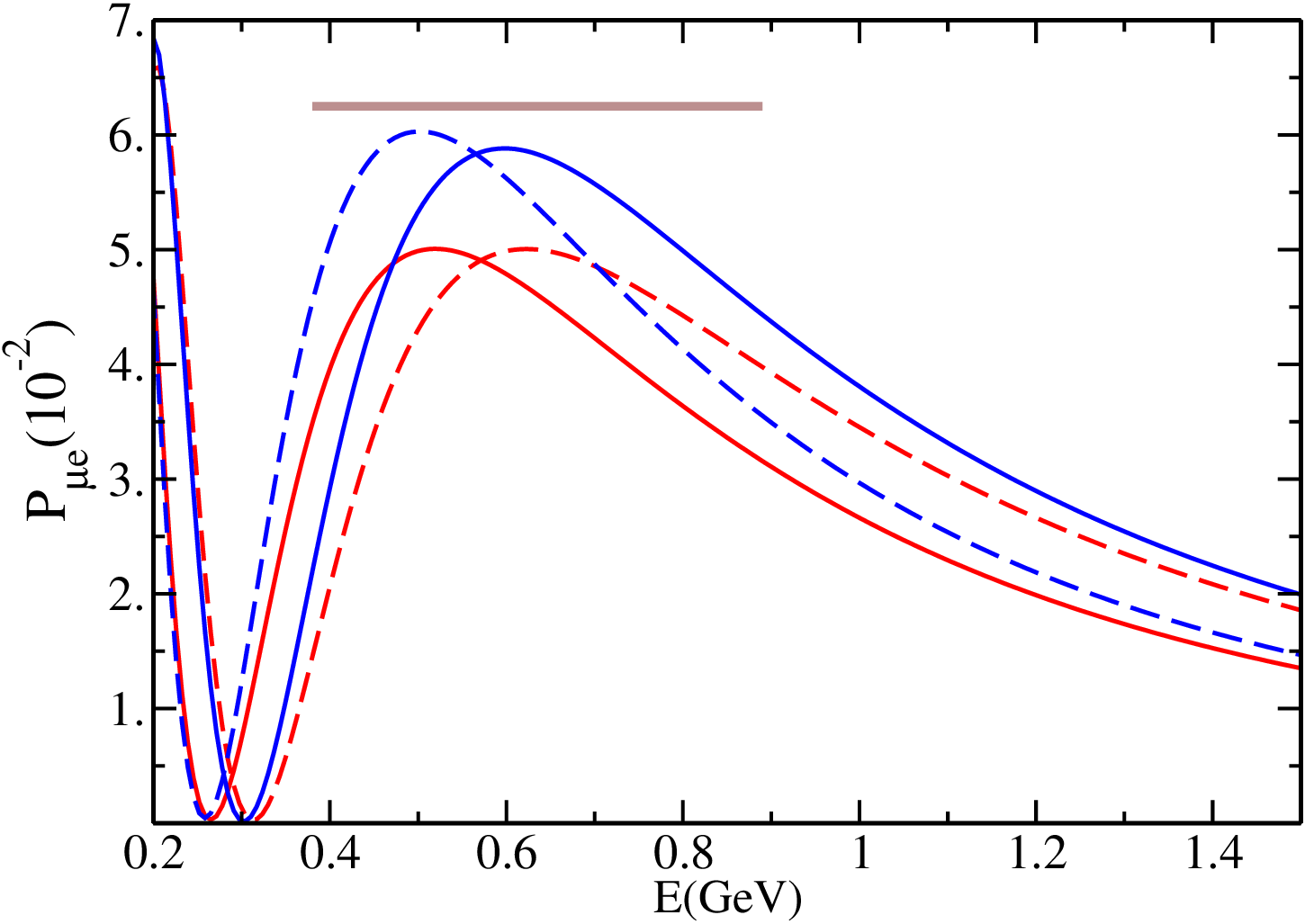}
\caption {${\mathcal P}_{\mu e}$  in matter versus neutrino energy for the T2K experiment. The blue curves depict the
normal hierarchy, red the inverse hierarchy. Solid curves depict positive $\theta_{13}$, dashed curves negative $\theta_{13}$ }
\label{fig2}
\end{center}
\end{figure}

\section{Symmetry}

Extracting neutrino mixing parameters from data is complicated by the existence of symmetries that lead to
degeneracies \cite{Bar02}, {\it i.e.} different mixing parameters yield the same oscillation probabilities. 
The MINOS \cite{Ada11b}, T2K \cite{Abe11}, and the future NO$\nu$A \cite{Now12} experiments are at an $L/E$ where such a
broken symmetry exists. 
In the limit of $\theta_{13}=0$ and no matter effects, these experiments are insensitive to the hierarchy. The symmetry pertinent 
to these experiments is the simultaneous interchange of hierarchy and the sign of $\theta_{13}$, 
giving a four-fold degeneracy. In Fig.~\ref{fig1} we depict the vacuum oscillation probability ${\mathcal P}_{\mu e}$ for the T2K experiment for 
full three-neutrino mixing and non-zero $\theta_{13}$. The four-fold degeneracy is partially broken by non-zero 
$\theta_{13}$. The solid curve represents both the 
normal hierarchy, positive $\theta_{13}$ as well as the inverse hierarchy, negative $\theta_{13}$ results while the 
dashed curve represents the inverse hierarchy, positive $\theta_{13}$ and the normal hierarchy, negative          
$\theta_{13}$ results. Non-zero $\theta_{13}$ breaks the four fold degeneracy down to two two-fold degeneracies.     

\begin{figure}[t]
\begin{center}
\includegraphics*[width=3.5in]{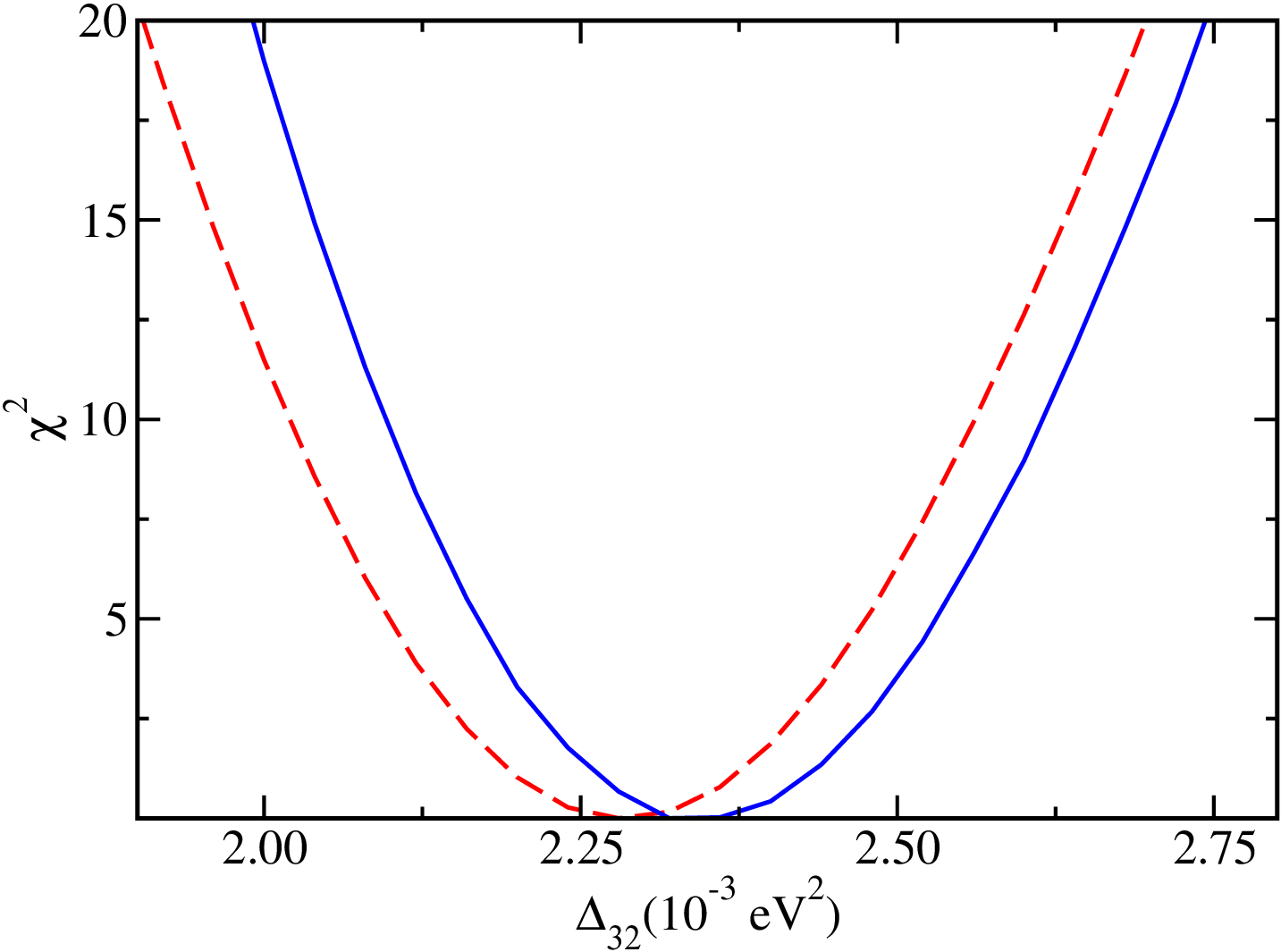}
\caption
{$\chi^2$ versus the mass-squared difference $\Delta_{32}$. The solid blue curve is the result using all the experiments
mention in the text. The red dashed curve excludes two of these experiments, MINOS anti-neutrino disappearance and T2K neutrino
disappearance.}
\label{fig3}
\end{center}
\end{figure}

We present, in Fig.~\ref{fig2}, ${\mathcal P}_{\mu e}$ versus neutrino energy including the MSW matter effects for 
the T2K experiment\cite{Abe11}. The matter interaction of the neutrino breaks the two fold symmetries by altering the magnitude 
of the oscillations while leaving the position of the peaks nearly unchanged. 
Spectral information that measures the position of the peaks maintains approximately the two-fold degeneracy of the vacuum oscillations. 
The magnitude of the signal then breaks this symmetry to distinguish the hierarchy. The combination of T2K and NO$\nu$A will be particularly powerful as they have vastly
different baselines and thus different matter effects. This will be helpful in disentangling the hierarchy, the
sign of $\theta_{13}$, and CP violation.

\begin{figure}[t]
\begin{center}
\includegraphics*[width=3.5in]{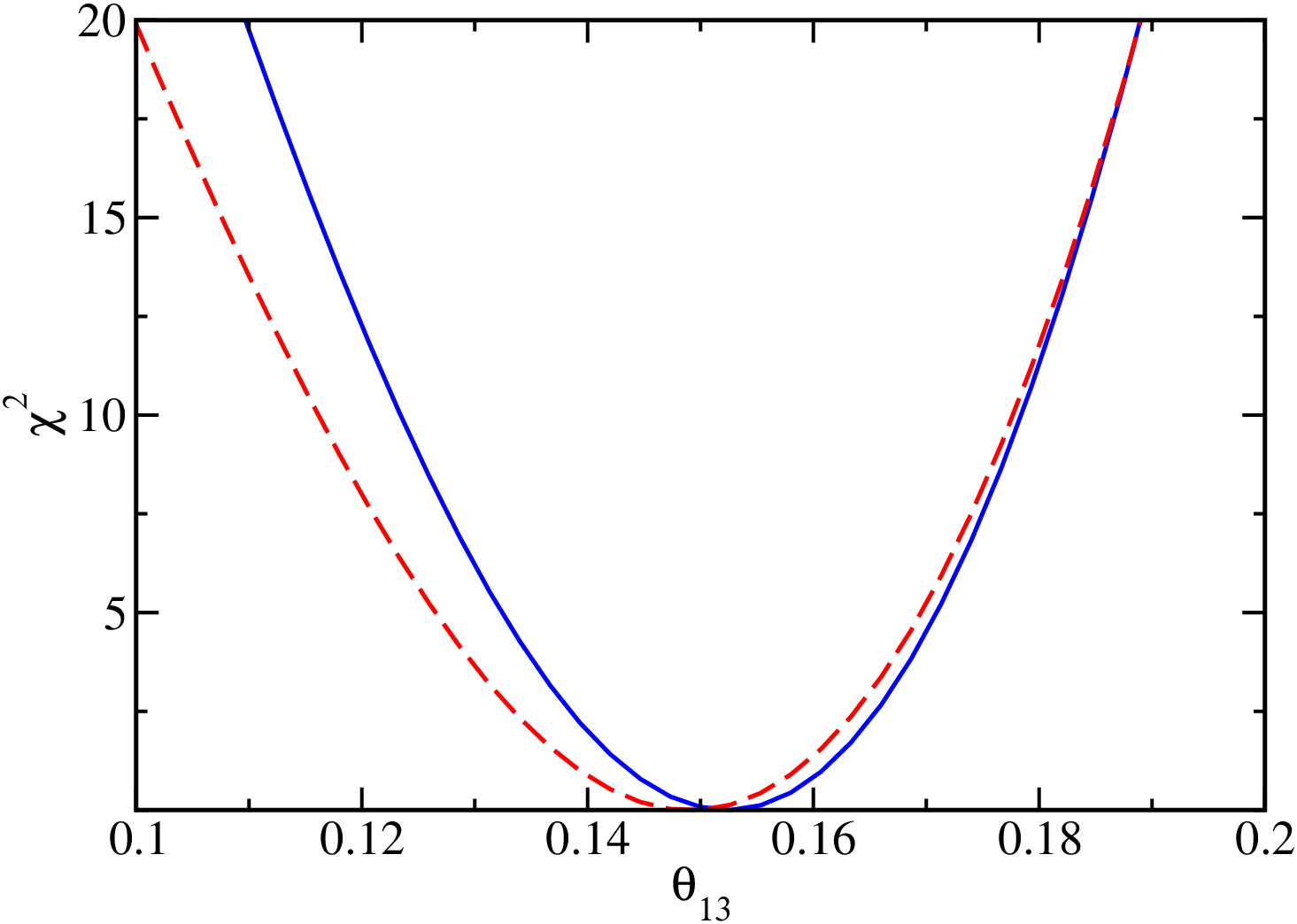}
\caption
{$\chi^2$ versus $\theta_{13}$.The solid blue curve is the result using all the experiments
mention in the text. The red dashed curve excludes RENO.}
\label{fig4}
\end{center}
\end{figure}

\section{Parameters}

In this section we present the results of a global analysis that includes the following experiments: 1) The
long-baseline muon disappearance experiments, MINOS neutrino\cite{Ada11}, MINOS anti-neutrino\cite{Ada12} and
T2K\cite{Abe12b}. 2) The one kilometer baseline reactor experiments, Daya Bay\cite{An12} and RENO\cite{Ahn12}. 3) The constraints on
$\theta_{23}$ from Super-K\cite{Hos06} as taken from the analysis of Ref.~\refcite{Roa09}.

In Fig.~\ref{fig3} we present the results of our analysis for  $\Delta_{32}$. Throughout
this work we calculate $\chi^2(\theta_{13},\theta_{23}, \Delta_{32})$ and marginalize over the non-displayed parameters. The solid blue curve
is the full analysis. The long-baseline MINOS neutrino disappearance\cite{Ada12} experiment dominates this results. The
dashed red curve is the result excluding the
MINOS anti-neutrino\cite{Ada11} and T2K neutrino\cite{Abe12b} experiments. We see they move the minimum upwards.
Our results are $\Delta_{32} = 2.45\pm 0.14 \times 10^{-3}$ eV$^2$, errors are the 90\% errors.

In Fig.~\ref{fig4} we present the results of our global analysis for the mixing angle $\theta_{13}$. The solid blue
curve is the full analysis. Daya Bay dominates this result. The red dashed curve is the full
analysis minus RENO. It is seen that RENO reduces the low error bar. Our results are $\theta_{13} = 0.152\pm 0.014$.

\begin{figure}[t]
\begin{center}
\includegraphics*[width=3.5in]{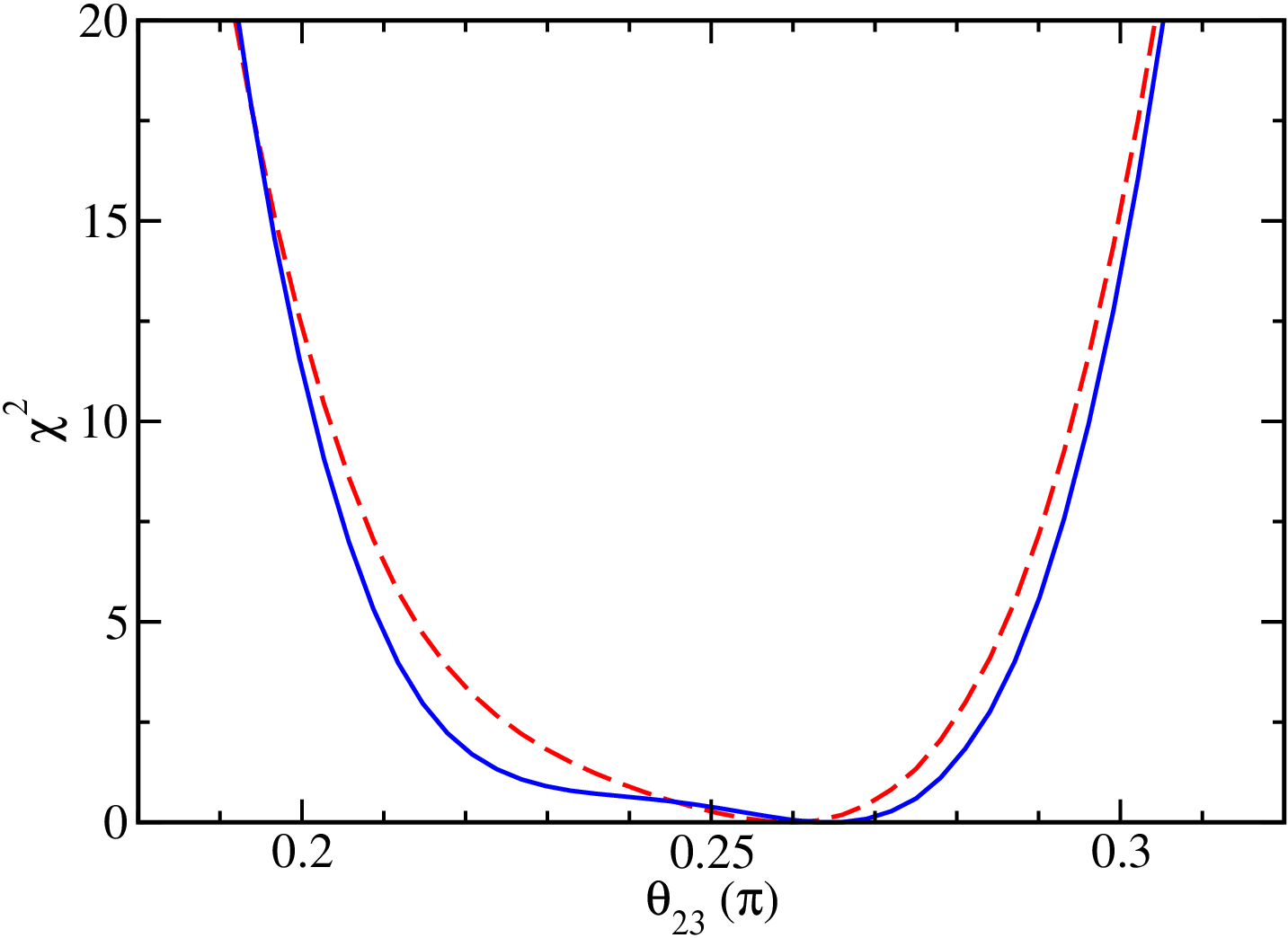}
\caption
{$\chi^2$ versus $\theta_{23}$. The solid blue curve is the result using all the experiments
mention in the text. The red dashed curve excludes the three long-baseline muon disappearance experiments that also
restrict the value of $\theta_{23}$.}
\label{fig5}
\end{center}
\end{figure}

In Fig.~\ref{fig5} we present the result of our global analysis for the mixing angle $\theta_{23}$. The solid blue
curve is the full analysis. The Super-K solar results dominate this result. The red dashed curve is the full
analysis minus the three long-baseline muon disappearance
experiments. We see that they have little effect. Our result remains the same as
Ref.~\refcite{Roa09}, $\theta_{23}=0.251^{+0.03}_{-0.05}~\pi$.

\section{Hierarchy}

\begin{figure}[b]
\begin{center}
\includegraphics*[width=3.5in]{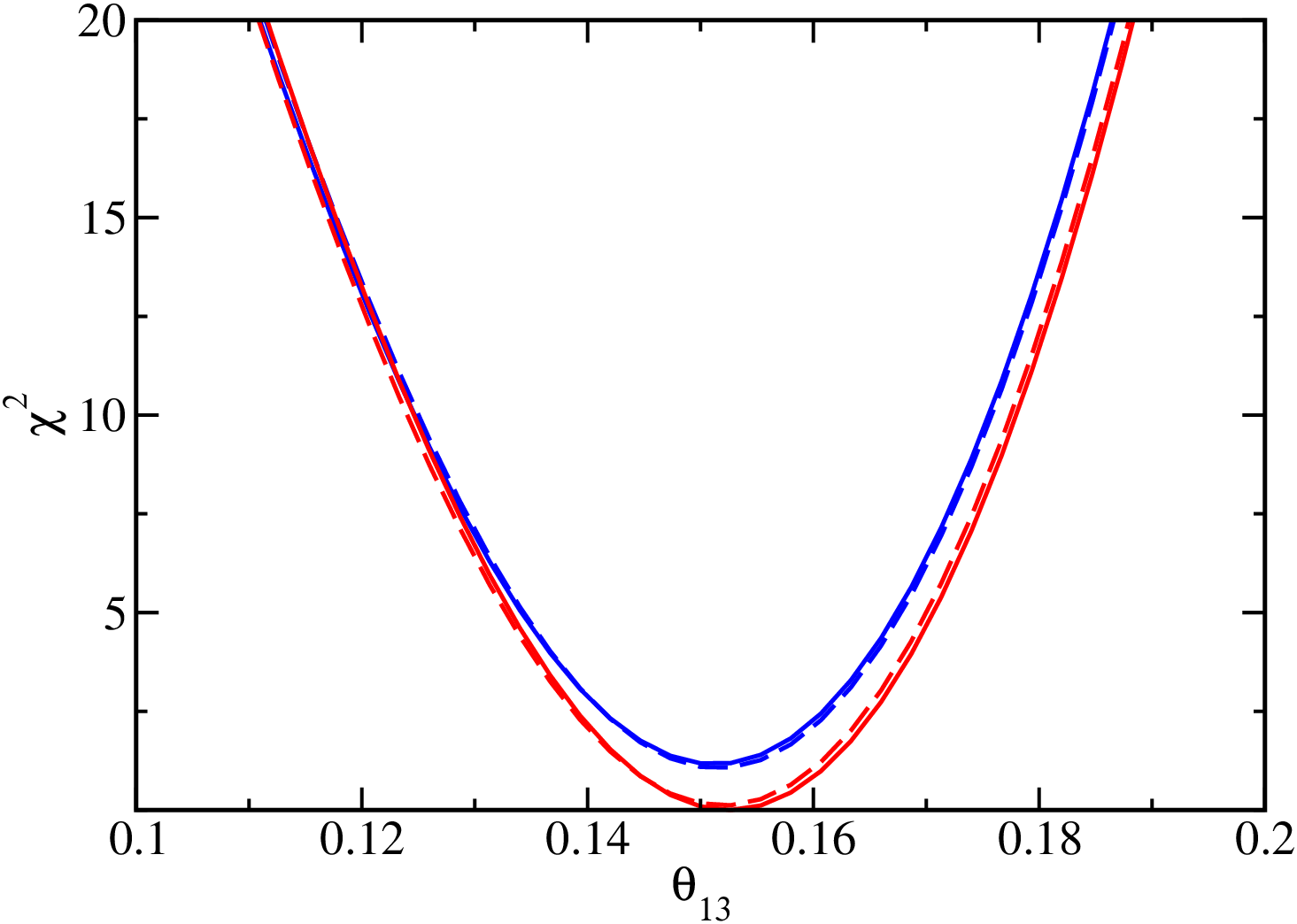}
\caption
{$\chi^2$ versus $\theta_{13}$ for all of the experiments as explained in the text plus the two long-baseline muon
appearance experiments, MINOS and T2K. The blue curves are the normal hierarchy, the red the inverse hierarchy. The
solid curves are positive $\theta_{13}$, the dashed negative $\theta_{13}$.}
\label{fig6}
\end{center}
\end{figure}

Two long-baseline experiments, the MINOS and T2K $\nu_\mu\rightarrow\nu_e$
experiments, were specifically designed to be at an $L/E$ where ${\cal P}_{\mu e}$ is sensitive to
the hierarchy, the sign of $\theta_{13}$, and CP violation. MINOS has taken its final data\cite{Ada13}. We have not completed 
an analysis of this data and here
present results from earlier data\cite{Ada11b}. This data was not taken with an off-axis beam
causing the background to be much larger than the signal. Also, the value of $L/E$ for MINOS does not place it at the
peaks seen in Figs.~\ref{fig1}, \ref{fig2}. It is sensitive to the hierarchy only. The data\cite{Abe11} 
from T2K contain only six counts. We add these two data sets to the data utilized in the previous section and present the results in
Fig.~\ref{fig6}. The blue curves depict the normal hierarchy, red the inverse hierarchy. Solid curves represent positive
$\theta_{13}$, dashed curves negative $\theta_{13}$. There is only sensitivity to the hierarchy, with the inverse
hierarchy preferred at about one sigma. Since T2K is presently running and NO$\nu$A is just starting, the future
looks quite promising.

\begin{figure}[t]
\begin{center}
\includegraphics*[width=3.5in]{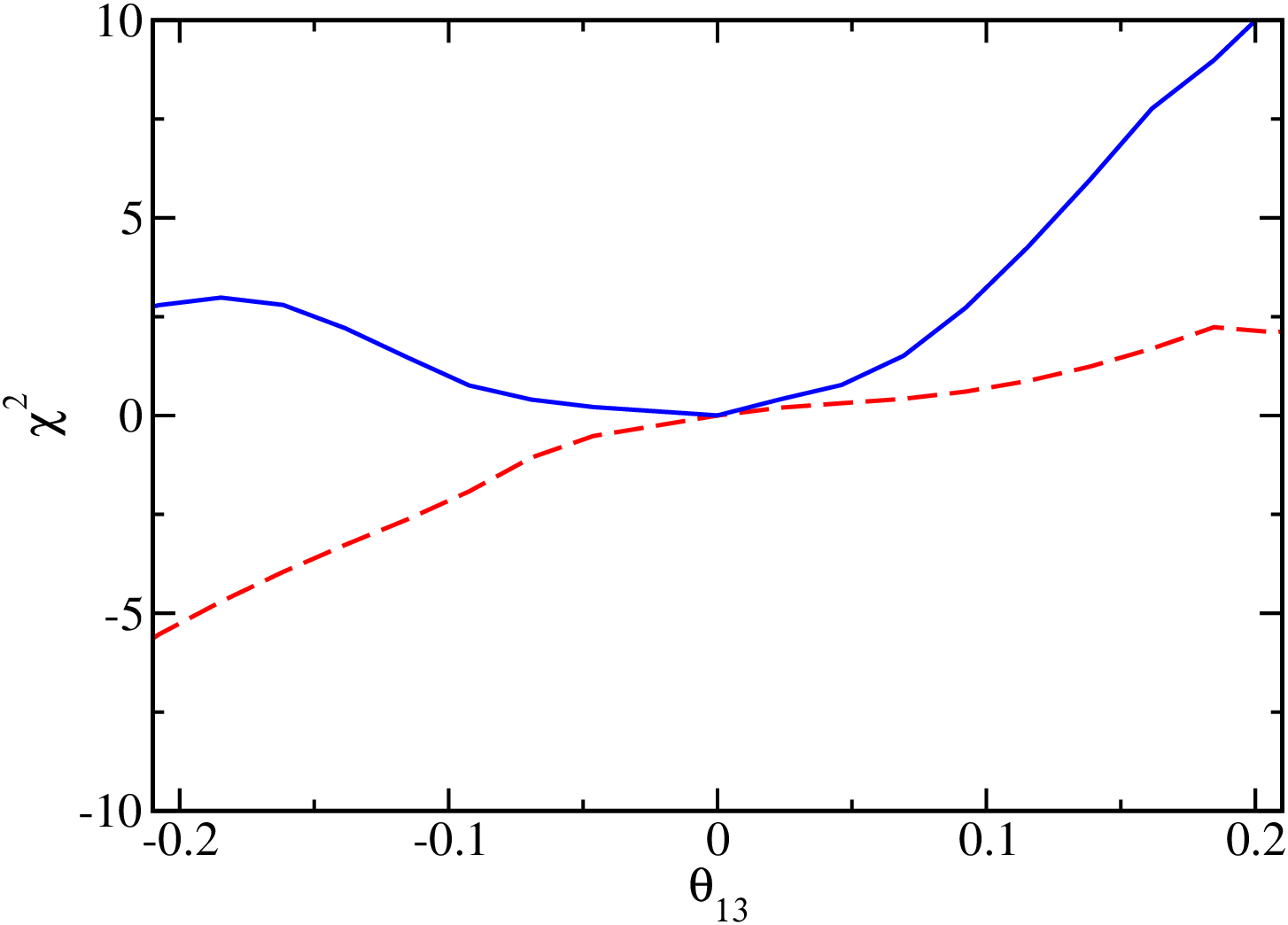}
\caption
{$\chi^2$ versus $\theta_{13}$ for Super-K atmospheric data. The solid blue curve is for the normal hierarchy, the dashed
red the inverse hierarchy.}
\label{fig7}
\end{center}
\end{figure}

Atmospheric data covers an extremely large range of $L/E$ values including those that discriminate between the
hierarchies, the value and sign of $\theta_{13}$, and the amount of CP violation.  In Fig.~\ref{fig7} we present $\chi^2$
versus $\theta_{13}$ for Super-K
as taken from Ref.~\refcite{Roa09}.
The blue curve is the
normal hierarchy while the red curve is the inverse hierarchy. The inverse hierarchy is preferred over the normal 
hierarchy. The source of this is the {\it lack} of excess 
electron neutrinos in the energy region from 3 to 7 GeV. This is the region where there are MSW matter resonances for
the normal hierarchy but not for the inverse hierarchy. The lack of an excess of electron neutrinos implies no resonances
and hence favor the inverse hierarchy. The effect is quadratic in $\theta_{13}$. The statistics at these
high energies are presently not so good. With the now known large value of $\theta_{13}$, should the lack of excess
electron neutrinos persist in future data, the atmospheric data alone could prove that the inverse hierarchy is the correct choice.

The linear terms in $\theta_{13}$ occur when there is interference between the two mass-squared differences. This
extends from the $L/E$ of T2K and NO$\nu$A of 590 km/GeV to where the effect is maximum\cite{Lat07} at about $1.5 \times
10^4$ km/GeV. For atmospheric data this occurs for data near the largest $L$ and the smallest $E$. Note that the slope of
the linear terms (the inverse hierarchy is dominated by the linear terms) also favors inverse hierarchy, plus favor
negative $\theta_{13}$ 

\begin{figure}[t]
\begin{center}
\includegraphics*[width=3.5in]{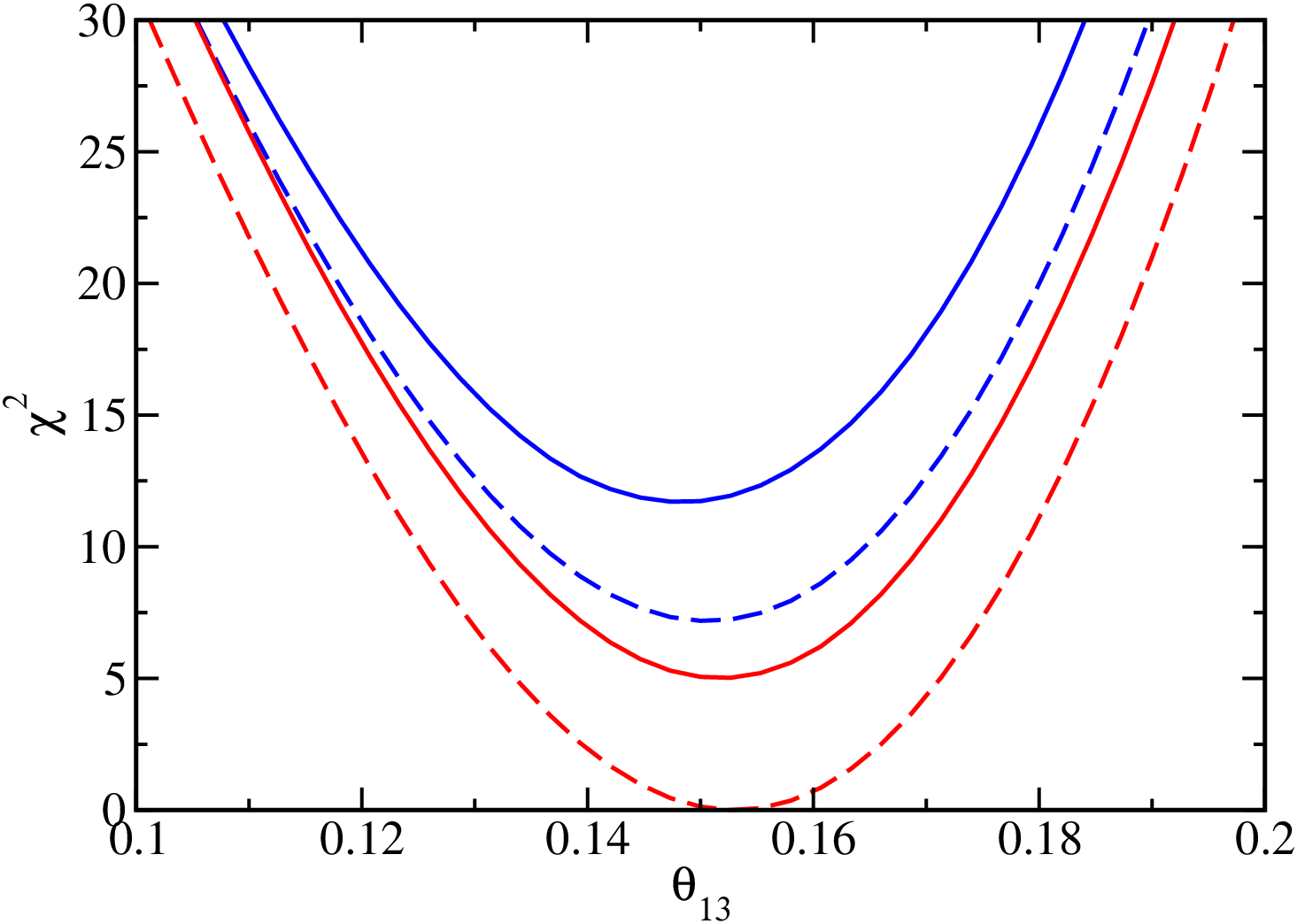}
\caption
{$\chi^2$ versus $\theta_{13}$ for the four solutinos corresponding to hierarchy and the sign of $\theta_{13}$. The blue curves  depict the normal 
hierarchy, the red curves the inverse hierarchy. The solid curves depict positive $\theta_{13}$, the dashed curves
 negative $\theta_{13}$.}
\label{fig8}
\end{center}
\end{figure}

We, unlike others, find four distinct and isolated minima, one for each value of the hierarchy and the sign of
$\theta_{13}$ as implied by the symmetry. In Fig.~\ref{fig8} we present the values of $\chi^2$ versus
$\theta_{13}$ for the four cases. We find that the negative hierarchy is favored over the positive hierarchy and that
negative $\theta_{13}$ is favored over positive $\theta_{13}$. The probability that each of the four probabilities is
correct is given in Table I. We use the Bayesian method proposed in Ref.~\refcite{Bur12} to obtain these probabilities.

\begin{table}[b]
\tbl{The best fit values of $\theta_{13}$ with their 90\% errors for the four combinations of hierarchy and the sign of
$\theta_{13}$. The probability that each combination is the correct one is also given}
{\begin{tabular}{lcccccc}
\toprule
hierarchy&sign $\theta_{13}$&$\theta_{13}$&\% probable\\
\colrule
normal&$+$&$0.148\pm0.015$&~0.3 \%\\
normal&$-$&$0.151\pm0.013$&~2.5 \%\\
inverse&$+$&$0.152\pm0.013$&~7.2 \%\\
inverse&$-$&$0.153\pm0.014$&90.0 \%\\
\botrule
\label{tab1}
\end{tabular}}
\end{table}

\section{Conclusions}

We have investigated the implications of the recent knowledge of the value of $\theta_{13}$ on
the determination of the physical hierarchy and the related question of the sign of $\theta_{13}$.
We find that the inverse hierarchy is preferred at the 97.2 \% level, while negative $\theta_{13}$ is preferred at the
92.5 \% level. The sensitivity that leads to these conclusions arises from the Super-K atmospheric data.

We are in disagreement with the results from Refs.~\refcite{For12} and \refcite{Fog12}, which differ from each other. We can only speculate on the
source of the differences. That we find four distinct minima in the $\chi^2$-space as implied by the symmetry while
others do not is encouraging for us. This could be caused by the use of the mass-squared dominance approximation that, as
we have pointed out\cite{Lat05b,Lat05c,Lat07}, does not converge to the correct answer in precisely the region where 
there is interference between the two mass-squared
difference oscillations. Alternatively, the second minimum for each hierarchy could have accidentally been overlooked. The preference for
the inverse hierarchy comes from the {\it lack} of excess electron neutrinos in the region of the MSW matter resonances.
A lack of excess electron neutrinos implies a lack of such resonances which favors the inverse hierarchy, something that
has been known for some time. 

We previously presented\cite{Ern12} an even more preliminary result that is consistent with the present results.
The upgraded work here incorporates the latest Daya Bay data, and more importantly marginalizes over parameters rather than fixing
them. We still need to update the two MINOS experiments to include data that came out after this talk. Most importantly, we are reviewing
our atmospheric analysis. It is this data that produce the different results for each of the four solutions. We will also
add CP violation.

\end{document}